\documentclass[%
 aip,
 amsmath,amssymb,
 reprint,%
]{revtex4-2}

\usepackage[T1]{fontenc} 
\usepackage{hyperref}
\usepackage{graphicx}
\usepackage{amsmath}
\usepackage{bm}
\graphicspath{{figures/}}
\usepackage{tikz}
\usepackage{xcolor}

\usepackage[normalem]{ulem} 

\begin{document}

\author{Bernadette Mohr}
\email{b.j.mohr@uva.nl}
\author{Diego van der Mast}
\affiliation{Van 't Hoff Institute for Molecular Sciences and Informatics Institute,
University of Amsterdam, Amsterdam 1098 XH, The Netherlands.}
\author{Tristan Bereau}
\affiliation{Van 't Hoff Institute for Molecular Sciences and Informatics Institute,
University of Amsterdam, Amsterdam 1098 XH, The Netherlands.}
\affiliation{Max Planck Institute for Polymer Research, 55128 Mainz, Germany}

\title{Condensed-phase molecular representation to link structure and
thermodynamics in molecular dynamics}


\begin{abstract}
Molecular design requires systematic and broadly applicable methods to extract
structure--property relationships. The focus of this study is on learning
thermodynamic properties from molecular-liquid simulations. The methodology
relies on an atomic representation originally developed for electronic
properties: the Spectrum of London and Axilrod-Teller-Muto representation
(SLATM). SLATM's expansion in one-, two-, and three-body interactions makes it
amenable to probing structural ordering in molecular liquids. We show that such
representation encodes enough critical information to permit the learning of
thermodynamic properties via linear methods. We demonstrate our approach on the
preferential insertion of small solute molecules toward cardiolipin membranes
and monitor selectivity against a similar lipid. Our analysis reveals simple,
interpretable relationships between two- and three-body interactions and
selectivity, identifies key interactions to build optimal prototypical solutes,
and charts a two-dimensional projection that displays clearly separated basins.
The methodology is generally applicable to a variety of thermodynamic
properties.
\end{abstract}

\maketitle

\section{Introduction}
Computational molecular design is rapidly becoming one of the most
exciting fields of our time thanks to its impressive developments and
broad applicability.\cite{curtarolo2013high, ferguson2017machine,
sidky2020machine, gkeka2020machine, dijkstra2021predictive,
bereau2021computational} The idea is simple: identify molecules or
materials with desirable properties. In practice, solving the
underlying inverse design problem remains challenging, requiring
extensive computational resources combined with an approach that
exploits the underlying physics and chemistry at hand. Electronic
properties have spearheaded the movement: quantum-mechanical (QM)
calculations (e.g., density-functional theory) over large numbers of
molecules have been successfully used in the context of machine
learning (ML) to predict various properties with increasing accuracy
and generalization.\cite{von2020exploring} In no small part is this
success due to the development of molecular representations: they
exploit physical laws (e.g., $r^{-1}$ scaling for Coulombic
interactions) and account for symmetries via
invariances.\cite{musil2021physics, smith2023topological} In the present study
we focus on thermodynamic properties, in particular in the context of
condensed-phase liquids.

The same principles ought to hold when moving from electronic to
thermodynamic properties: molecular representations form the basic
ingredients to describe structural features, and any physical prior
will help learning performance. While electronic properties typically
focus on single molecules in the gas, our consideration of
thermodynamic properties brings two specificities:
\begin{enumerate}
  \item Thermodynamics underlines the role of conformational entropy.
  Beyond a static structure, the diversity of conformations heavily
  impacts the energetics, calling for phase-space (Boltzmann)
  averaging;
  \item The condensed phase involves a molecule embedded in a dense
  environment, highlighting the balance of covalent and non-covalent
  interactions.
\end{enumerate}
The question addressed by this study is how to efficiently learn
structure--property relationships from (bio)molecular simulations of
thermodynamic properties.

We take clues from the field of glassy dynamics. Impressive ML
developments have been made to establish new insight into the relevant
structure--dynamics relationships.\cite{cubuk2015identifying,
bapst2020unveiling} These remarkable strides have required large and
complex deep neural networks. However, significantly smaller ML models
can be used when exploiting relevant physics: representations that
focus on the local as well as neighboring
structure.\cite{boattini2021averaging} The representations often
consist of structural order parameters, in particular radial and
angular structure functions. While the radial (i.e., two-body)
component measures the density of particles, akin to a radial
distribution function (RDF), the angular terms are inspired by
bond-orientational order parameters, i.e., three-body
interactions.\cite{steinhardt1983bond} The description of molecular
systems in terms of increasing number of interacting particles is
called a \emph{body-order} expansion.\cite{musil2021physics}

\begin{figure*}[htbp]
  \includegraphics[width=0.7\linewidth]{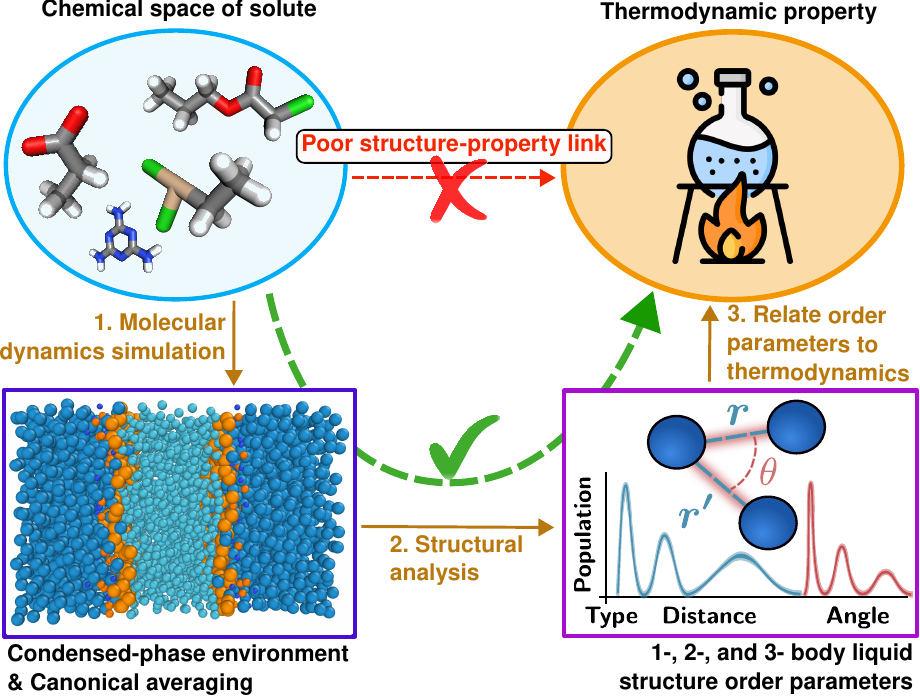}
  \caption{This study considers the identification of
    structure--property relationships between a solute molecule and a
    target thermodynamic property. Rather than directly learning the
    relationship, we propose a three-step process: (1) molecular
    dynamics simulations of the solute in its condensed-phase
    environment; (2) structural analysis of the liquid structure; (3)
    relate structural order parameters and thermodynamic properties.
    The learning procedure thus relies on features that incorporate
    relevant physics, including Boltzmann phase-space averaging,
    liquid environment, and collective effects. Our methodology is
    able to identify complex structure--property relationships, even
    with a simple linear model.}
  \label{fig:pipeline}
\end{figure*}

In this work, we adapt the idea of structural body-order interactions
to learn thermodynamic properties in molecular simulations. We propose
to start from an atomic representation originally developed for the
machine learning of electronic properties: the Spectrum of London
Axilrod-Teller-Muto (SLATM).\cite{huang2018fundamentals,
huang2020quantum} SLATM provides a body-order expansion through a
histogram of one-, two-, and three-body atomic contributions.
Moreover, it does not distinguish between covalent and non-covalent
interactions, making it well suited for a condensed phase. Finally, we
extend its role to a Boltzmann ensemble by averaging over snapshots of
a molecular dynamics (MD) trajectory.\cite{rauer2020hydration,
weinreich2021machine, weinreich2022ab} Fig.~\ref{fig:pipeline}
sketches our approach: from chemical-space compound screening to
thermodynamic properties via the structural analysis of MD
simulations. When establishing structure--property relationships, we
expect the structural order parameters to encode critical information
that will ease the learning process.  

\begin{figure}[htbp]
  \includegraphics[width=0.95\linewidth]{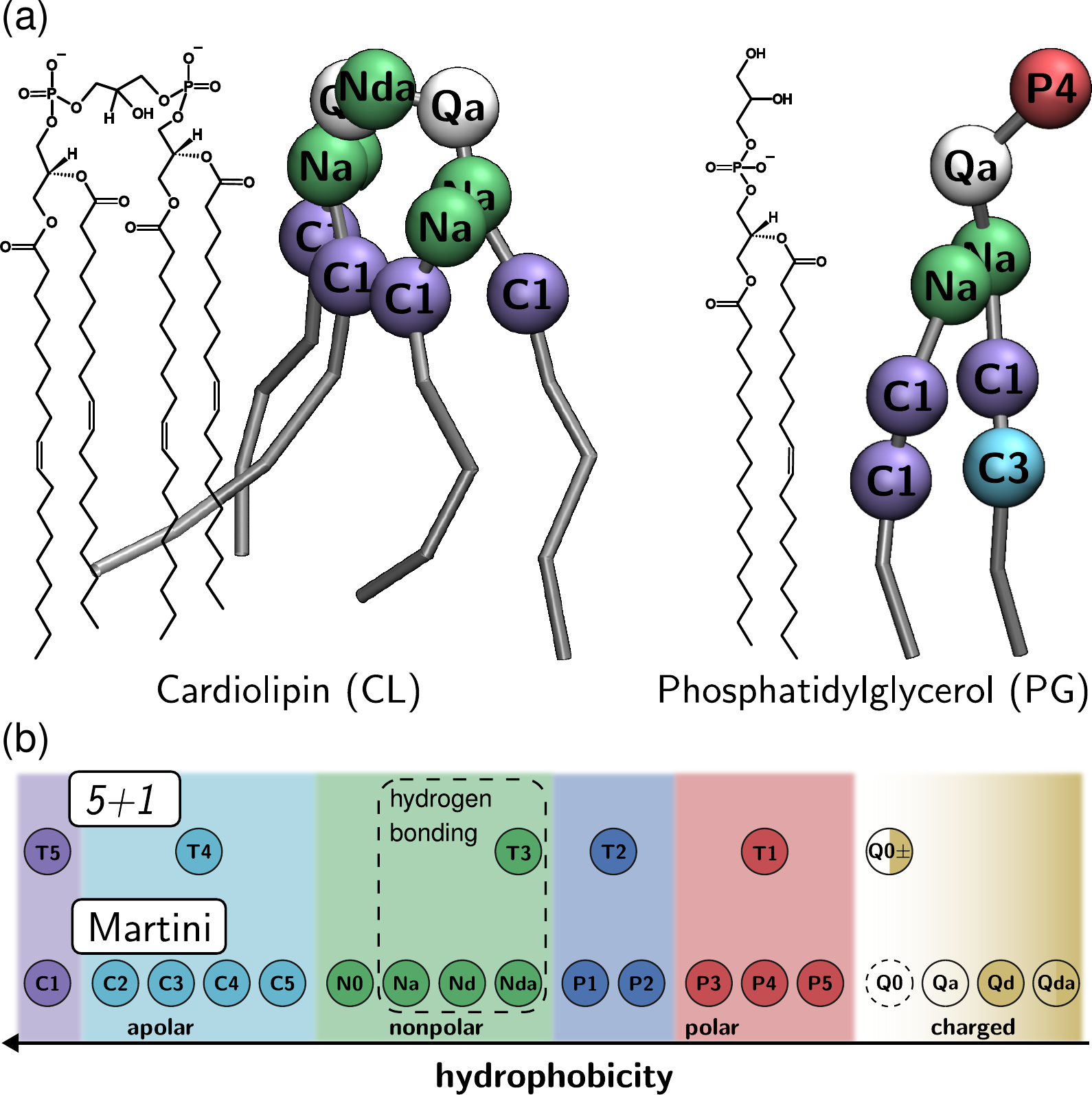}
  \caption{(a) Chemical structures of cardiolipin (CL) and
    phosphatidylglycerol (PG) next to their coarse-grained (CG)
    Martini representation. CG beads relevant for solute-lipid interactions
are placed according to the chemical structure. (b) CG bead types for Martini
and the reduced
    \emph{5+1} force field. The hydrophobicity scale summarizes the
    beads' physicochemical characteristics. Structures and cartoon
    representations were rendered using ChemSketch and
    VMD.\cite{chemsketch, HUMP96}}
  \label{fig:lipids}
\end{figure}

The application we focus on is a challenging biomolecular system:
Lipid selectivity of small molecules in mitochondrial membranes. The
problem involves the subtle identification of preferential
interactions between two similar lipids: cardiolipin (CL) and
phosphatidylglycerol (PG).\cite{dudek2017role, paradies2014functional,
elias2019curvature, houtkooper2008cardiolipin, pennington2019role,
paradies2019role, gonzalvez2013barth, yi2022effects}
Fig.~\ref{fig:lipids}a shows the chemical structures of CL and PG. The
binding selectivity of a small molecule between CL and PG membranes
amounts to a \emph{relative} free-energy difference, $\Delta \Delta
G$. Each free-energy difference quantifies the insertion of said
compound from bulk water to one membrane interface.
Fig.~\ref{fig:lipids}a highlights the chemical resemblance between one
CL molecule and a pair of PG lipids, emphasizing the difficulty of the
problem.

The complexity of the system compounded with the size of drug-like
chemical compound space makes an atomistic modeling approach
intractable. Instead, we base our investigation on coarse-grained (CG)
MD simulations. Coarse-graining averages over atomic degrees of
freedom to only describe larger super-particles or
\emph{beads}.\cite{voth2008coarse, noid2013perspective} Beyond the
computational appeal of faster MD simulations, a certain class of CG
models has the appealing property to reduce the size of
chemical-compound space.\cite{dobson2004chemical,
bereau2021computational} Compressing chemical space translates to a
more efficient compound screening---a valuable property to establish
structure--property relationships. CG models that can reduce the size
of chemical space have a top-down parametrization strategy: they aim
at modeling large-scale behavior by defining a finite set of bead
types, which encode specific physicochemical flavors. Critically, it
is the number of bead types that scales the (reduced) size of chemical
space of the CG model. While we base our study on the biomolecular CG
Martini model,\cite{marrink2007martini, alessandri2021martini} we will
use a further reduced yet compatible CG model, made of fewer bead
types, to efficiently screen for small
molecules.\cite{kanekal2019resolution} Fig.~\ref{fig:lipids}b
illustrates the reduction in number of bead types between the original
Martini and our reduced so-called \emph{5+1} force field. We
previously used this approach to devise a rigorous discovery pipeline
combining CG simulations, free-energy calculations, and active
learning, which, taken together, led to the identification of design
rules.\cite{mohr2022data} We further showed that such CG simulations
can be used to propose small-molecule probes for experimental
validation, with exciting results both \emph{in vitro} and \emph{in
vivo}.\cite{kleinwachter2022clib}

The CG simulations will be used as test system for the ensemble SLATM
approach. Our dataset consists of $n=439$ solute small molecules, for
which we have calculated the target property, $\Delta \Delta G$. Here
we will run a single isothermal-isobaric MD simulation per compound
and lipid environment, so as to compute the averaged structural order
parameter. Because the resulting molecular representation is difficult
to interpret, we will apply dimensionality reduction on the set of
$n=439$ order parameters. To demonstrate the effectiveness of our
representation, we will limit ourselves to a linear method: principal
component analysis (PCA). The linearity of the method will also be
leveraged through interpretability: we will extract the key two- and
three-body interactions that are most relevant to modulate
selectivity. The gathered insight will allow us to construct
prototypical solutes that optimize for the target property. Finally,
we will show that a two-dimensional projection in PCA coordinates
displays clearly separated basins of solutes with high and poor CL
selectivity, effectively generating a clear structure--property map.

\section{Methods}

In the following we cover the three methodological parts sketched in
Fig.~\ref{fig:pipeline}: ($i$) Molecular dynamics simulations; ($ii$)
Structural analysis; and ($iii$) Relating order parameters to
thermodynamics.

\subsection{Molecular dynamics simulations}

Coarse-grained (CG) molecular dynamics (MD) simulations were run using
GROMACS 2020 and Martini parameters tailored to GPU
acceleration.\cite{abraham2015gromacs, de2016martini} We used an
integration time step $\delta t = 0.02\,\tau$, where $\tau$ is the
natural unit of time of the model. The simulations were kept at
constant temperature ($T=300~\text{K}$) and pressure
($P=1~\text{bar}$) using the Langevin thermostat and Parrinello-Rahman
barostat.\cite{parrinello1981polymorphic} Electrostatic interactions
were calculated using particle-mesh Ewald
summation.\cite{darden1993particle}

Membranes were generated using the CHARMM-GUI Martini
maker.\cite{qi2015charmm} The cardiolipin (CL) and phosphatidylglycerol
(PG) membranes consist of 98 and 118 lipid molecules, respectively.
They were solvated in water, as well as sodium ions to maintain charge
neutrality. Bulk water systems consisted of 974 water beads, as well
as sodium and chloride ions to mimick the ion concentration of the
membrane systems. More details about the MD simulation setups and
parameters can be found in Mohr \emph{et al.}\cite{mohr2022data}

\subsubsection{Coarse-grained modeling}

All lipids, water, and ion particles were represented using the
standard CG Martini 2 force field with refined polarizable models for
water and ions.\cite{marrink2007martini, michalowsky2017refined,
michalowsky2018polarizable} For the solute compounds, we used a
\emph{reduced} and compatible Martini-like CG force
field.\cite{kanekal2019resolution} Compared to Martini's 14 bead
types, the reduced force field only defines 6: 5 neutral and one
charged, denoted \{\textsf{T1, T2, T3, T4, T5}\} and
\{\textsf{Q0$\pm$}\}, respectively (i.e., we define a single charged
bead type, though the charge can take a positive or negative sign). We
herein refer to the reduced model as the \emph{5+1} force field.
Fig.~\ref{fig:lipids}b highlights the placement of the fewer CG beads
on the hydrophobicity axis. Utilizing fewer bead types compresses the
size of chemical space, used here to more efficiently screen across
solutes.

Solute compounds were constructed by considering various graph
representations and a variety of CG bead types from the \emph{5+1}
force field.\cite{mohr2022data} We limited the number of beads to up
to five, in order to roughly stay within the molecular weight
prescribed in Lipinsky's rule of five for drug-likeness of small
molecules.\cite{lipinski2004lead} We applied angles and constraints to
the compound structures according to their geometry (Figure S2). This small
change in conditions compared to the previously preformed free-energy
calculations is warranted due to the dependence of the structural order
parameter on non--conflicting particle coordinates. See SI for more details
on the \emph{5+1} force field and solute graph representations.

The subsequent structural analysis of a solute in a membrane
environment will monitor CG beads from both force fields: 
\begin{itemize}
  \item \emph{All} beads from the reduced \emph{5+1} force field, so
  as to screen across the solute's chemical space, i.e.,
  \{\textsf{T1}, \textsf{T2}, \textsf{T3}, \textsf{T4}, \textsf{T5},
  $\text{\textsf{Q0}}\pm$\};
  \item Only \emph{some} beads from Martini: those involved in
  describing the CL and PG membrane environments, as well as the water
  and ion models, i.e., \{\textsf{Nda}, \textsf{P4}, \textsf{Qa},
  \textsf{Na}, \textsf{C1}, \textsf{C3}, \textsf{POL}, \textsf{PQd}\}.
\end{itemize}
The combination yields a set of $N=14$ different bead types, which
will impact the dimensionality of the structural order-parameter
vectors described below.

\subsubsection{Alchemical free-energy calculations of selectivity}

Our target thermodynamic property is the selectivity of a solute to
preferably insert in a CL membrane compared to a similar PG membrane.
Selectivity thus corresponds to a relative thermodynamic affinity
between the two membrane environments. We quantify the individual
insertions by means of transfer free energies from bulk water to the
interfacial region of the membrane bilayer, denoted 
\begin{equation}
  \Delta G_{\text W
  \rightarrow \text M} = \Delta G^{\text M} - \Delta G^{\text W}.
  \label{eq:transfer_free_energy}
\end{equation}
Accordingly, selectivity is measured by the difference of transfer
free energies between PG and CL environments
\begin{equation}\label{eq:selec}
  \Delta\Delta G = \Delta 
  G^\text{CL}_{\text{W}\rightarrow \text{M}} 
  - \Delta G^\text{PG}_{\text{W}\rightarrow \text{M}}.
 \end{equation}

Both terms in Eq.~\ref{eq:transfer_free_energy} were calculated using
\emph{relative} alchemical free-energy calculations: we focused on the
change in free energy when solvating the solute. The water environment
was simulated using a simple water box. The membrane simulation
consisted of an equilibrated lipid bilayer with added solute placed at
the interface, i.e., close to the lipid headgroups, which embodies the
main chemical difference between PG and CL. 

Alchemical free-energy calculations consisted of successive coupling
of all nonbonded interactions (i.e., van der Waals and electrostatics)
between a solute and its surrounding environment, together with the
use of soft-core potentials.\cite{torrie1977nonphysical,
chipot2007free, mey2020best} We applied 40 intermediate coupling steps
for each interaction type to ensure adequate sampling. We subsequently
estimated free energies using the MBAR method and the \texttt{pymbar}
package.\cite{shirts2008statistically, pymbar} For more details about
the free-energy calculations, see Mohr \emph{et
al.}\cite{mohr2022data} 

\subsubsection{Trajectory analysis}

The present structural analysis solely relies on the fully coupled
alchemical state of the system, while other states were entirely
discarded. For each one of the $N=439$ compounds, we ran and analyzed
an MD simulation of total simulation time $\Delta t = 20,000~\tau$, and
extracted 200 frames. For each snapshot, we centered the simulation around the
solute and kept information up to a radial distance of 1.1~nm. Trajectory
processing was performed using
\texttt{MDAnlysis}.\cite{oliver_beckstein-proc-scipy-2016,
michaud2011mdanalysis}

\subsection{Structural analysis}

\subsubsection{The Spectrum of London Axilrod-Teller-Muto
(SLATM) representation: atomic case}

\begin{figure}[htbp]
  \centering
  \includegraphics[width=0.9\linewidth, keepaspectratio]{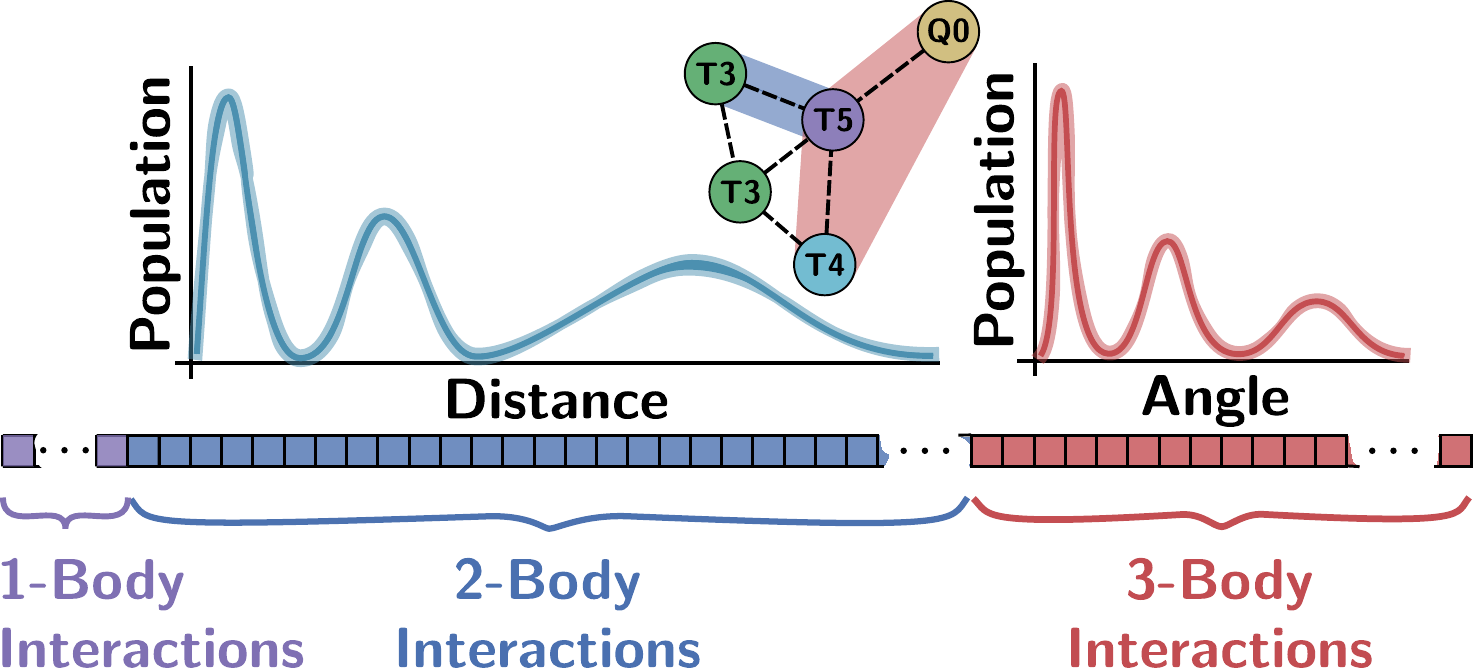}
  \caption{Schematic of a SLATM histogram with decomposition in one-, two-,
 and three- body contributions: Particle counts (purple), pairwise
 interactions (blue), and triplets (red). Inset: cartoon representation
 of interacting beads with example two- and three-body interactions
 around the \textsf{T5} particle. Dashed lines emphasize that
 interactions need not be covariant.}
  \label{fig:slatm}
 \end{figure}

The Spectrum of London Axilrod-Teller-Muto (SLATM) representation
describes an atomic environment as a vector of one-, two-, and
three-body interactions occurring within a cutoff (Figure
\ref{fig:slatm}).\cite{huang2018fundamentals, huang2020quantum} SLATM
ignores the notion of covalent bonding. The representation features
translational, rotational, and permutation invariance. Given a
particle $i$ (atom or CG bead), let $I$ refer to its atom or bead
type---one out of $N$ types defined by the force field. We denote by
$\bm{x}_i$ the SLATM representation of particle $i$, as a sum over
body-order contributions
\begin{enumerate}
\item The one-body term, $x_i^{(1)}$, simply accounts for the identity
of the particle, denoted $Z_I$---the elemental atomic number for
particle $i$ in an atomistic representation. For the present CG
resolution, we remedy the lack of elemental number by assigning $Z_I$
an arbitrary (but unique) value;
\item The two-body interaction, $x_{i,J}^{(2)}(r)$, represents the
population of pairwise interactions between $i$ and all other
particles of type $J$, as a function of radial distance, $r$;
\item  The three-body bond-angle interaction,
$x_{i,JK}^{(3)}(\theta)$, describes the interactions between $i$ and
all other particles of types $J$ and $K$, as a function of the angle,
$\theta$, and averaged across interparticle distances.
\end{enumerate}

The radial and angular dependence of the two- and three-body
interactions are binned along their respective intervals: $[0,
r_\textrm{cutoff}]$ and $[-\frac \pi 9, \frac{10\pi} 9]$. For ease of
notation we represent the binned interaction as a vector. For
instance, the two-body representation between particle $i$ and all
others of type $J$ yields
\begin{equation}
  \bm x_{i,J}^{(2)} =
    \begin{bmatrix}
      x_{i,J}^{(2)}(r_0)\\
      \vdots\\
      x_{i,J}^{(2)}(r_{N_{\textrm{b}}-1}^{(2)})
    \end{bmatrix},
\end{equation}
where $r$ denotes the interparticle distance, and the size of the
vector is given by the number of radial histogram bins, $N_b^{(2)}$. A
similar representation is considered for three-body interactions
between particle $i$ together with all combinations of types $J$ with
$K$, $\bm x_{i,JK}^{(3)}$, which would bin over the angle between a
triplet of particles. SLATM then concatenates over all possible
pairwise and triplet types to yield
\begin{equation}
  \bm{x}_i = \Biggl[
    \underbrace{x_i^{(1)}}_{\text{one-body}},
    \underbrace{\bm x_{i,J_0}^{(2)},
    \ldots, \bm x_{i,J_N}^{(2)}}_{\text{two-body}},
    \underbrace{\bm x_{i,J_0K_0}^{(3)}, \ldots, \bm x_{i,J_NK_N}^{(3)}}
    _{\text{three-body}}
  \Biggr]^\intercal.
  \label{eq:slatm_vector}
\end{equation}

Functional forms for two- and three-body interactions follow the
London dispersion forces and the Axilrod-Teller-Muto
potential.\cite{london1930theorie, axilrod1943interaction,
muto1943force} The body-order interactions read
\begin{align}
  x^{(1)}_i &= Z_I \\
  x^{(2)}_{i,J}(r) &= \frac 12 Z_I Z_J \sum_{j \in J} 
  g(r-R_{ij})\frac 1{r^6} \\
  x^{(3)}_{i,JK}(\theta) &= \frac 13 Z_I Z_J Z_K \sum_{j \neq i} \sum_{k \neq j \neq i}  
  g(\theta - \theta_{ijk}) \nonumber \\
  &\quad \times \frac{1 + \cos\theta \cos\theta_{jki} \cos\theta_{kij}}
  {(R_{ij}R_{ik}R_{kj})^3},
\end{align}
where $R_{ij}$ and $\theta_{ijk}$ are the pairwise distance and
triplet angle, respectively, and two- and three-body interactions are
smoothened by a Gaussian function
\begin{equation}
  g(x) = \frac 1{\sigma\sqrt{2\pi}} 
  \exp\left(- \frac{x^2}{2\sigma^2} \right).
\end{equation}

We used the implementation of Christensen \emph{et al.} and adapted
some of their parameters for use with CG
resolution.\cite{christensen2017qml} Notably, the widths of the
Gaussian kernels were set to $\sigma = 0.3~\textrm{\AA}$ and
0.2~radian for distances and angles, respectively. The bin widths of
the histograms were set to $0.2~\textrm{\AA}$ and $0.2$~radian, and
the radial cutoff to $r_\textrm{cutoff} = 8.0~\textrm{\AA}$.

\subsubsection{Boltzmann-ensemble averaging}

A single configuration is not statistically significant, i.e., we
require a Boltzmann average of the representation, $\langle \bm
x_i\rangle$. By ergodicity, we approximate the Boltzmann-ensemble
average by a time average, i.e., over snapshots of the MD trajectory.
Though we gather 500 equidistant frames along the trajectory, we only
keep a subset of 200 to exclude those whose solute lies further away
from the target depth of insertion (i.e., at the lipid headgroup
interface). We calculate the Boltzmann-averages over these 200 snapshots of
each atomic SLATM.

\subsubsection{Molecular SLATM}

Rather than focusing on atomic representations, we describe the
behavior of an entire molecule at once. To do so, we sum over all CG
beads of a molecule of interest, $\mathcal{M}$, so as to yield the
Boltzmann-averaged molecular representation
\begin{equation}
  \langle\bm{\mathcal{X}}\rangle = \sum_{i \in \mathcal{M}} 
  \langle \bm x_i\rangle. \label{eq:mol_slatm}
\end{equation}
The sum in Eq.~\ref{eq:mol_slatm} requires a separation of
contributions in the various bead types. A single atomic SLATM
contains 1, $N$, and $N(N+1)/2$ one-, two-, and three-body terms.
When summing over multiple particles, the molecular SLATM will feature $N$,
$N(N+1)/2$, and a subset of $N^3$ types, pairs, and triplets, respectively. For
the number of triplets considered, see SI Sec.~2. In this study, we consider
$N=14$ bead types, leading to 14, 105, and 1361 contributions.

Each two- and three-body contribution has high dimensionality, it is a
histogram over a range of distances or angles. To reduce the
dimensionality of the molecular SLATM vector, we averaged over the
distance and angular information. Averages were normalized against the
sum over the corresponding distances or angles (Equation S1).

\subsection{Relating order parameters to thermodynamics}

Recall that our target thermodynamic property is a solute's
selectivity to CL and against PG
membranes. The following describes our tailoring of the molecular
SLATM representation to focus on selectivity, and the use of principal
component analysis (PCA) to establish a structure--property relation.

\subsubsection{Tailoring SLATM to membrane selectivity}

The power-law behavior used for the two- and three-body interactions
lead to strong heterogeneities in the SLATM bins. Order-of-magnitude
differences are commonly observed, making their immediate use for any
ML analysis potentially difficult. Instead here we work with the
logarithm of the molecular SLATM, so as to compress the space. 

Focusing on the \emph{difference} in observed interactions between CL
and PG environments, our quantity of interest is the difference
between the two log-transformed molecular representations, leading to
\begin{equation}\label{eq:slatms}
	\Delta\langle\bm{\mathcal{X}}\rangle =
\ln\frac{\langle\bm{\mathcal{X}}\rangle_{\text{CL}}}
{\langle\bm{\mathcal{X}}\rangle_{\text{PG}}}.
\end{equation}
For each one of the $n=439$ herein considered compounds, we computed
the structural order-parameter vector,
$\Delta\langle\bm{\mathcal{X}}\rangle$. Further details are included in the
SI.

\subsubsection{Principal component analysis (PCA)}

Each structural order-parameter vector, $\Delta
\langle\bm{\mathcal{X}}\rangle$, is of high dimensionality: $D=1480$. To reduce
the dimensionality and effectively tease out the contributions most
relevant to thermodynamic selectivity, we apply a simple methodology:
principal component analysis (PCA).\cite{pearson1901liii,
hotelling1933analysis, wold1987principal, jolliffe2002principal} PCA
looks for a set of orthogonal directions that maximizes the variance
of the zero-mean data matrix, $\bm{\hat X}$, of dimension $n \times
D$, by solving the eigenproblem
\begin{equation}
  \textrm{Cov}(\bm{\hat X}, \bm{\hat X}) \bm v_k = \lambda_k \bm v_k,
\end{equation}
where $\lambda_k$ and $\bm v_k$ are the $k$-th eigenvalue and
unit-norm eigenvector, respectively. Similarly, the linear combination
$\bm{\hat X} \bm v_k$ is called the $k$-th principal component
(PC)---a scaled eigenvector. The elements of the eigenvectors $\bm
v_k$ are called the \emph{PC loadings}.\cite{jolliffe2016principal}
Intuitively, eigenvectors indicate the directions of high variance in
a set of samples, while eigenvalues represent the corresponding
amount, via the variance of the PCs. The proportion of variance
explained up to dimension $d$ is given by $\sum_{i<d}\lambda_i /
\sum_{j<D}\lambda_j$.\cite{glielmo2021unsupervised}

The PCA representation then consists of choosing a number of
components $d$ (where, typically, $d \ll D$), and projecting the
original data onto the eigenvectors as $\bm Y = \bm X \bm V$, where
$\bm V$ is a matrix of dimension $D \times d$ containing the first $d$
eigenvectors. Correlating lower-dimensional PCs to target properties
offers strong interpretability, thanks to the possibility to transform
back from PCs to original coordinates.\cite{ferguson2017machine,
glielmo2021unsupervised} 

We used the PCA implementation of the \texttt{scikit-learn} package
with the random seed set to a constant value for
reproducibility.\cite{tipping1999mixtures, bishop2016pattern,
scikit-learn} We performed no whitening of the data. For computational
efficiency, we used the PCA module using randomized singular value
decomposition, utilizing appropriate dimensionality and shape of the
SLATM arrays.

\subsubsection{PCA of molecular SLATM vectors depends almost
exclusively on two- and three-body interactions}

Though in principle all three bodies of interaction play a role in the
PCA analysis of $\Delta \langle \bm{\mathcal{X}} \rangle$
(Eq.~\ref{eq:slatms}), the one-body contributions are virtually
negligible. Indeed, the two lipid environments display almost the same
collections of bead types. The headgroup beads, \textsf{Nda} and
\textsf{P4}, are the distinguishing characteristics between CL and PG,
respectively (see Figure \ref{fig:lipids}). This difference is
systematically present in all $\Delta \langle \bm{\mathcal{X}}
\rangle$. Consequently, PCA places minimal importance on the one-body
contributions relative to the higher-order interactions. In the
following, we thus limit our evaluation to the two- and three-body
interactions.

\subsubsection{Physicochemical interpretation of the principal
components}

Interpretation of the main PCs was achieved by cross-correlation with
several (physicochemical) descriptors. All descriptors are normalized
by the number of CG beads in the solute, to account for the
heterogeneity in solute sizes. The descriptors include the
Water--octanol partitioning of the solutes, $\Delta G_{\text W
\rightarrow \text{Ol}}$ (see SI Sec.~1.1); Number of solute polar
beads, i.e., \textsf{T1} and \textsf{T2}; Number of solute charged
beads, i.e., \textsf{Q0}; Number of solute beads that offer
hydrogen-bond-like characteristics, i.e., \textsf{T3}; and The $l^2$
norm of the structural order-parameter vector,
$|\Delta\langle\bm{\mathcal X}\rangle|_2$. We relied on linear
regression to measure the correlation, quantified by the coefficient
of determination, $R^2$.\cite{camacho2010data}

\section{Results and Discussion}

The following describes the results of the methodology sketched in
Fig.~\ref{fig:pipeline} in the context of small solute molecules
interacting with either cardiolipin (CL) or phosphatidylglycerol (PG)
membranes. We run MD simulations, extract structural order-parameter
vectors (here in the form of the molecular SLATM), and subsequently
analyze them using a principal component analysis (PCA). We first
relate some of the first principal components (PCs) to physicochemical
properties. We then focus on the PC most relevant for selectivity, and
identify key two- and three-body interactions. Finally, we establish
linear structure--property relationships between PCs and selectivity
for CL membrane.

\subsection{Physicochemical interpretation of PCA
eigenvectors}\label{sec:crosscorr}

The amount of variance explained by the eigenvalues ideally prescribes
a number of PCs to retain $d \ll D$. Upon inspection, we find no clear
change in regime, but rather a smooth behavior (Fig.~S5). We focus
here on the first six eigenvectors, representing 77\% of the overall
variance. 

\begin{figure}[htbp]
  \includegraphics[width=0.8\linewidth]{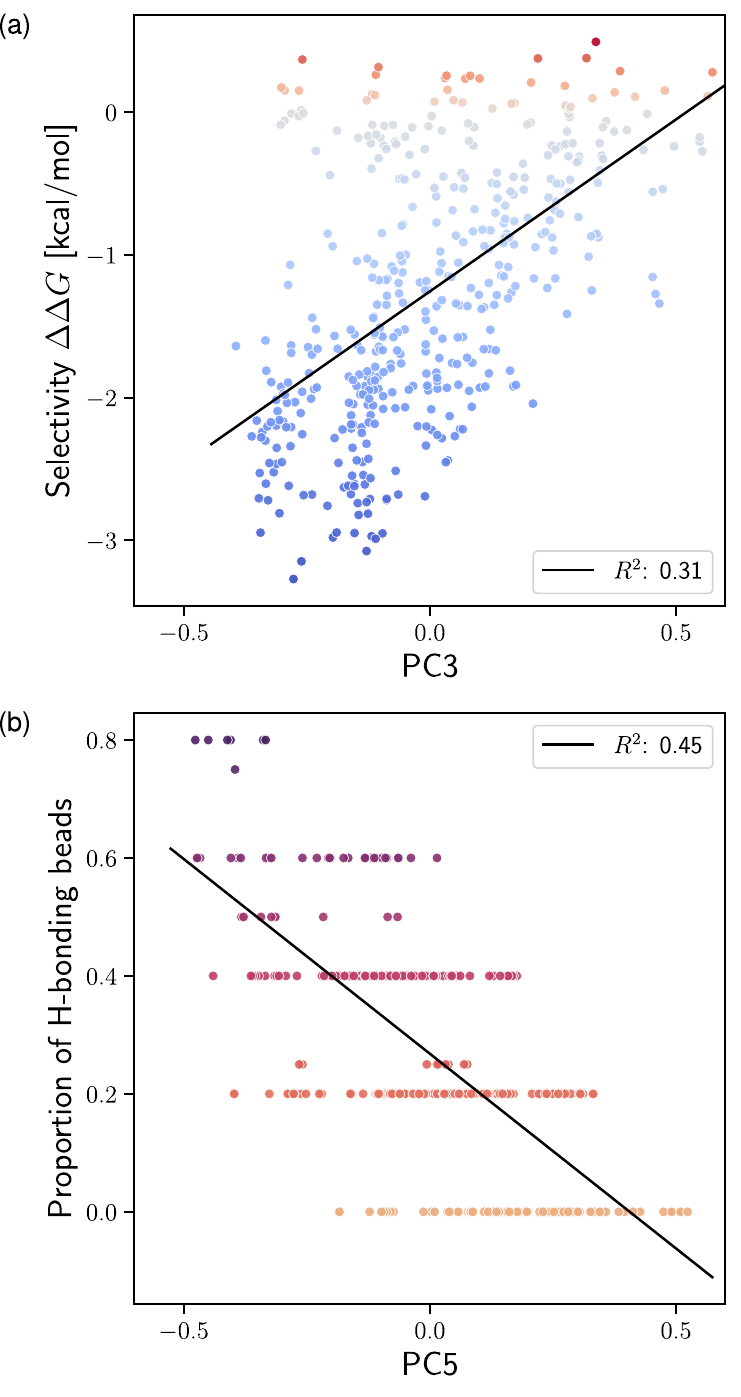}
  \caption{Cross-correlation of (a) the third principal component
 (PC3) to CL selectivity $\Delta\Delta G$, and (b) PC5 to the ratio of
 hydrogen-bonding beads in each solute. The color gradients further
 visualize the respective physicochemical descriptor represented on
 the vertical axis. The lines represent best fits from linear
 regression.}
  \label{fig:cc_pc3one}
\end{figure}

To interpret the first six PCs, we cross-correlate them with different
physicochemical descriptors. Fig.~\ref{fig:cc_pc3one}a shows the
correlation between the third component, PC3, against selectivity
itself, $\Delta \Delta G$. We measure a meaningful coefficient of
determination $R^2=0.31$, while cross-correlation with the other main
PCs yields virtually 0 (Fig.~S6). It is not surprising to find
correlation between the PCs and the target property, because of our
construction of the structural order-parameter vector. Indeed,
$\Delta\langle\bm{\mathcal{X}}\rangle$ focuses on the
\emph{difference} in observed interactions of a solute between the two
lipid environments. Though expected, the lack of correlation with any
other main PCs makes for a clear map between solute and target
selectivity, via a \emph{single} PC.

Simultaneously, we find that PC3 also correlates strongly with other
physicochemical descriptors: water--octanol partitioning free energy,
$\Delta G_{\text{W}\rightarrow\text{Ol}}$ ($R^2=0.36$); polar bead
types, \textsf{T1} and \textsf{T2} ($R^2=0.57$); and charged bead
types, \textsf{Q0} ($R^2=0.47$). However, PC3 does not correlate
significantly with bead types \textsf{T3}, associated with a CG proxy
for hydrogen-bonding (Figs.~S8--S10). Taken together, the direct
association of PC3 to selectivity hints at the role played by $\Delta
G_{\text{W}\rightarrow\text{Ol}}$, polar beads, and charged beads in
modulating CL selectivity. In fact, $\Delta
G_{\text{W}\rightarrow\text{Ol}}$ is a key quantity in the
parametrization of CG Martini, and in particular that of the reduced
\emph{5+1} force field.\cite{kanekal2019resolution} The design rules inferred
from our previous active-learning study similarly highlighted the effects of
polar and charged beads.\cite{mohr2022data}

Other PCs also exhibit some physicochemical interpretation, as shown
in Figs.~S6--S11. We find that PC1 and PC2 weakly correlate with polar
beads ($R^2=0.14$) and hydrogen-bonding beads ($R^2=0.13$),
respectively. The other main PCs correlate more significantly to
physicochemical descriptors: PC4 associates with both water--octanol
partitioning ($R^2=0.22$) and charged beads ($R^2=0.11$). PC5 strongly
correlates with \textsf{T3} types associated with hydrogen-bonding
($R^2=0.45$, Fig.~\ref{fig:cc_pc3one}b), and to a smaller extent with
the norm of the structural order-parameter vector ($R^2=0.32$) as well
as the number of charged beads ($R^2=0.21$). Finally, PC6 almost
exclusively and strongly correlates with the norm of the structural
order-parameter vector ($R^2=0.51$). It is not clear to us whether
this mirrors a sensitivity to an overall difference between CL and PG
environments, or whether the metric is biased by particular
coordinates of $\Delta\langle\bm{\mathcal{X}}\rangle$. 

Overall, the relatively straightforward association of PCs to few
physicochemical descriptors likely arises from the CG resolution of
the model itself, which reduces the number of relevant degrees of
freedom. In addition, we point at the effective role played by our MD
structural order-parameter vectors, which exacerbates the relationship
between salient features of the solute in its condensed-phase
environment with the target thermodynamic property.

\subsection{Identification of key interactions to design selective
solutes}\label{sec:structure}

Correlation of relevant PCs to selectivity is only a means to an end.
What we care to understand is the role played by specific (two- and
three-body) interactions in modulating selectivity. Fortunately, the
linearity of PCA allows us to easily transform back to the space of
$\Delta\langle\bm{\mathcal{X}}\rangle$ and read off the contribution
of every single interaction. To this end, we focus on the scaled PC
loadings (Equ. (S2)), i.e., the elements of the PCA eigenvectors. The
scaled PC loadings for the various PCs down to an absolute value of 1.0
are reported in Figs.~S12 and S13. However, not all components carry
equal importance. Recall from Fig.~\ref{fig:cc_pc3one}a that PC3
correlated positively with selectivity. On the other hand, strong
selectivity values tend to be large and negative (i.e., they are
free-energy differences). Given the positive correlation between PC3
and $\Delta\Delta G$, we expect PCs with \emph{negative} coefficients
to contribute to strong selectivity.

\begin{figure*}[htbp]
  \centering
  \includegraphics[width=0.8\linewidth, keepaspectratio]{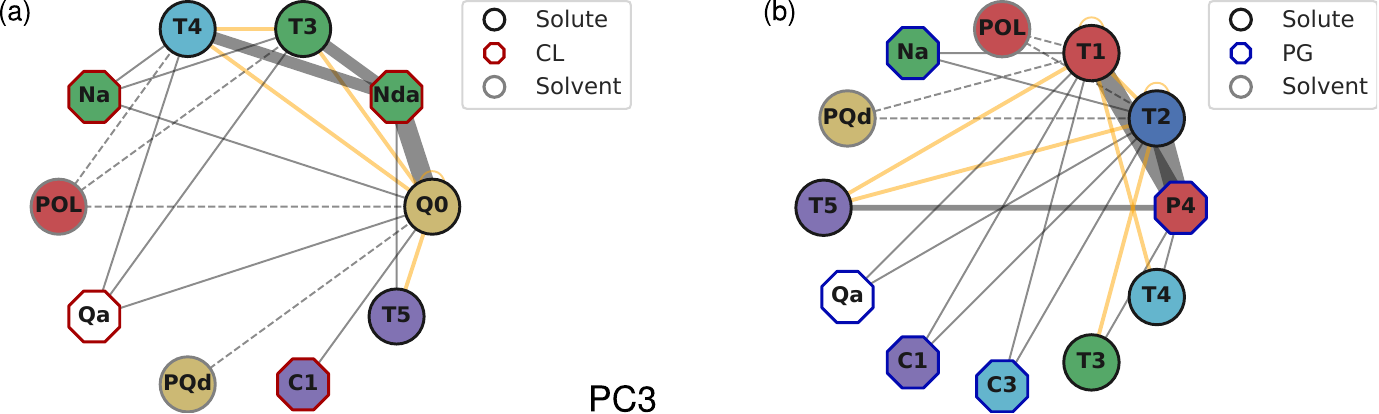}
  \caption{Graph of the interactions of PC3 with dominant negative PC
    loadings for the (a) CL and (b) PG systems. The edge width scales
    with the occurrence of the interaction. Orange edges show
    interactions between solute beads, and dashed edges represent
    interactions with beads used to model water or sodium ions.}
  \label{fig:graphs}
\end{figure*}

Fig.~\ref{fig:graphs} reports the interactions that have negative PC3
eigenvector components (see SI for the other PCs). The interactions
are displayed on a graph, where nodes and edges correspond to bead
types and interactions, respectively. Because edges are inherently
pairwise, three-body interactions are projected down onto the relevant
pairwise counterparts. Both bead types and interactions have specific
visual features depending on the system: solute, lipid, or solvent.
Importantly, the thickness of the edges emphasizes the occurrence of a bead
pair in the dominant scaled PC loadings, and thus the relevance of the
interaction. Panels a and b display the CL and PG systems, respectively.
First, they highlight the central role played by the \textsf{Nda} and
\textsf{P4}
lipid beads. We recall from Fig.~\ref{fig:lipids}a that these are the
two bead types that specifically distinguish CL from ($2\times$) PG.
For CL, \textsf{Nda} predominantly interacts with \textsf{Q0},
\textsf{T3}, and \textsf{T4}. For PG, \textsf{P4} interacts primarily
with \textsf{T1} and \textsf{T2}. For both membranes, these
contributions largely reflect the strengths of two-body interactions.
In addition, they are further reflected in many of the relevant
three-body interactions, sometimes accompanied by other bead types:
\textsf{T5} for CL; \textsf{T3}, \textsf{T4}, and \textsf{T5} for PG.
Furthermore, the role of the solvent is highlighted via key
interactions with the \textsf{POL} and \textsf{PQd} bead types.

\begin{figure}[htbp]
  \centering
  \includegraphics[width=0.9\linewidth]{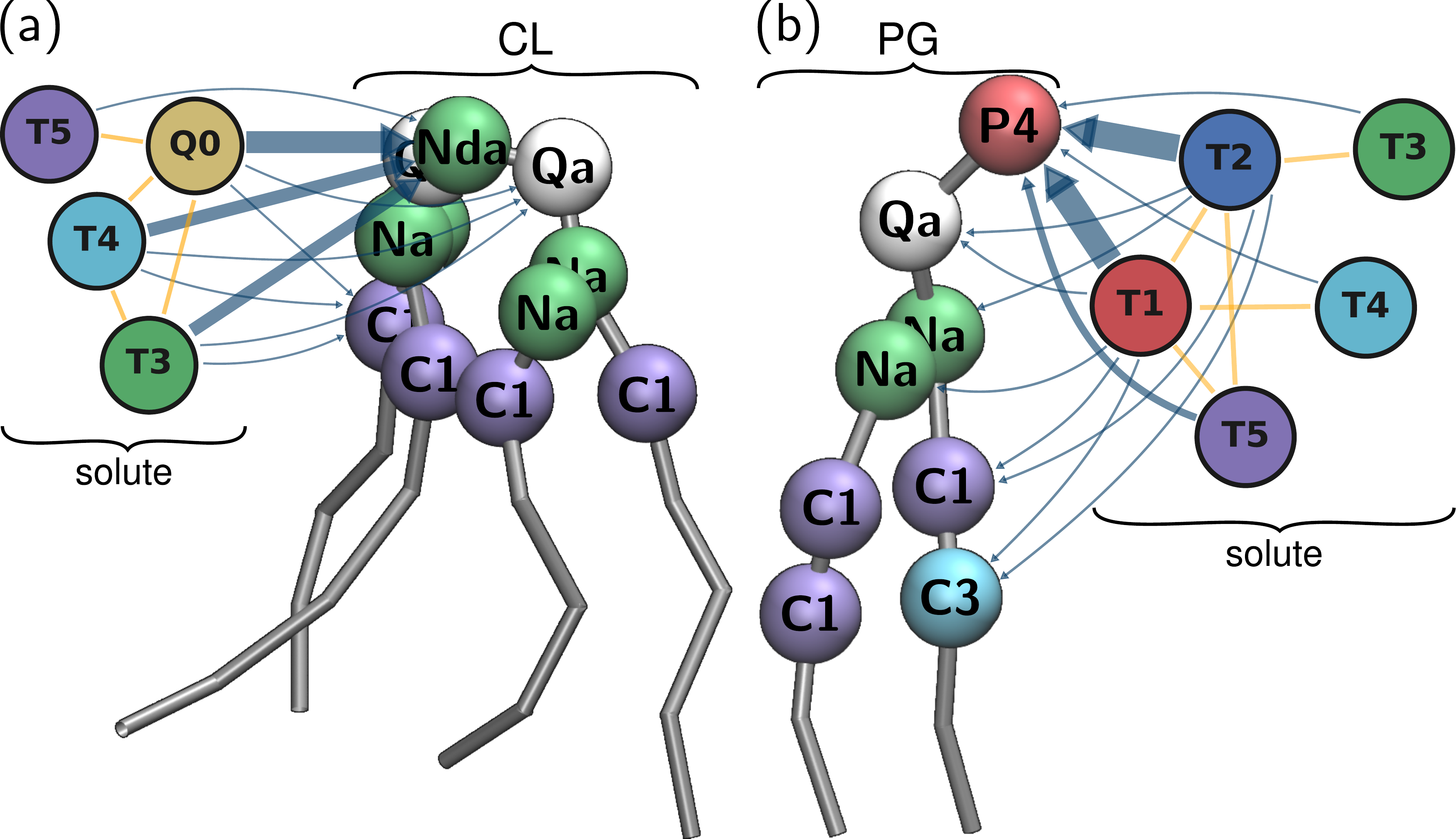}
  \caption{Three-dimensional illustration of the dominant interactions
    of PC3 reported in Fig.~\ref{fig:graphs} for (a) CL and (b) PG.
    The sets of beads form hypothetical solutes that would favorably
    interact with either lipid.}
  \label{fig:3d_illu}
\end{figure}
 
Leveraging information from this analysis further, we can visualize
favorable geometric arrangements of beads to enhance selectivity.
Fig.~\ref{fig:3d_illu} reconstructs information from the interaction
graphs to place prototypical solutes around the two lipids. Solute
beads are placed manually around the lipids so as to illustrate the
information of Fig.~\ref{fig:graphs}. The arrow widths further reflect
the interaction strengths, mirroring the PC loadings. The figure
emphasizes the role played by some of the bead types, and clearly
conveys the idea that different bead types will favorably associate
with either CL or PG. Panel a, which targets CL, better illustrates
relevant solute characteristics for the target property at hand in
this work.

\subsection{Charting selectivity in low-dimensional maps}

\begin{figure*}[htbp]
  \includegraphics[width=\linewidth, keepaspectratio]{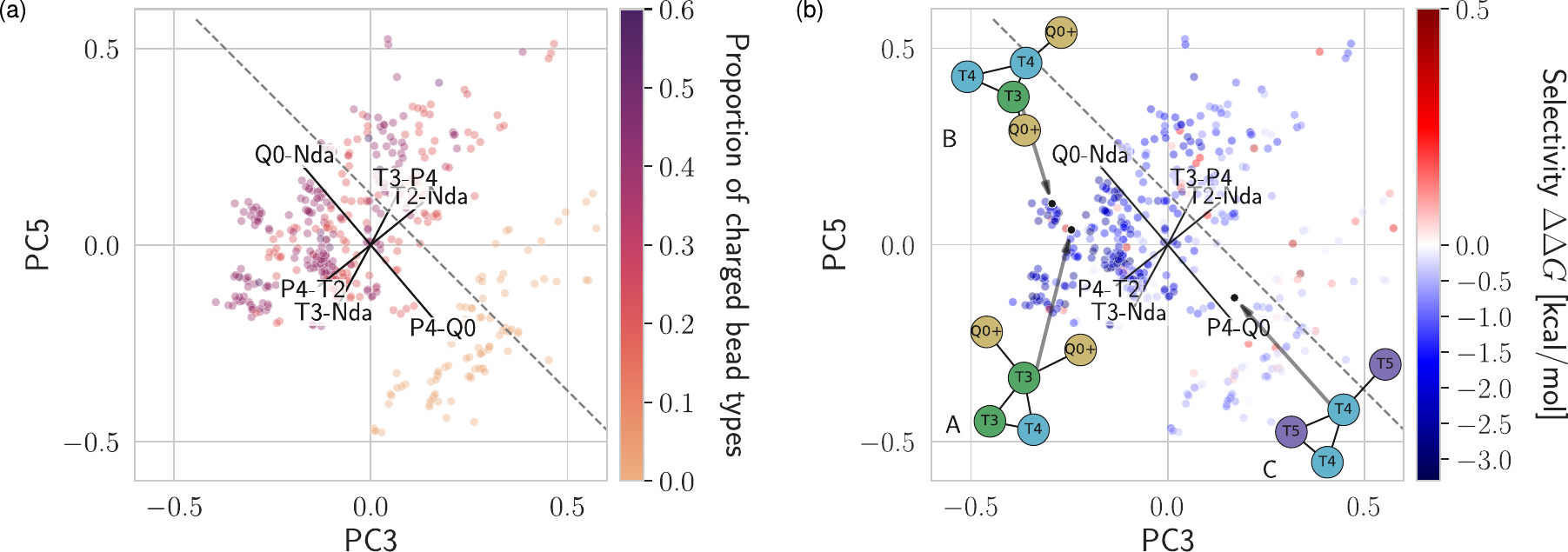}
  \caption{Biplots of PC3 and PC5, colored by (a) the ratio of charged
    beads per solute and (b) and (c) selectivity, $\Delta\Delta G$.
    The values of the principal components are scaled to the interval
    $[-1, 1]$. In (a) and (b), we show the six highest eigenvector
    coefficients of two-body interactions. In (b), three example
    compounds are classified for CL selectivity by PCA, without
    calculating their respective partitioning free-energy difference
    $\Delta\Delta G$.}
  \label{fig:prediction}
\end{figure*}

Beyond the relationship between individual PCs and selectivity, we
look for more insight by combining pairs of components. We iterate
through all pairs of PC1--6, each time generating a two-dimensional
map or embedding, populating it with the $n=439$ solutes based on
their PCA coordinates, and coloring the points according to the
different physicochemical descriptors (Figs.~S21--S32). Out of all
combinations, the pair PC3--PC5 stands out in its high overall
correlation to several descriptors, including selectivity.  The
two-dimensional map is reproduced in Fig.~\ref{fig:prediction}a.  The
combination is somewhat expected: the high correlation of PC3 and PC5 alone
was already reported in Fig.~\ref{fig:cc_pc3one}a and b, respectively.
Fig.~\ref{fig:prediction}a shows that the combination of PC3--PC5
creates two clear basins in terms of proportion of charge in the
solute. Remarkably, this projection simultaneously leads to a
\emph{separation} between poorly and highly selective solute
compounds, as evidenced in Fig.~\ref{fig:prediction}b. This separation
is clearly visible between the upper-left and lower-right corners of
the space. The basin of high selectivity associates with low and high
values of PC3 and PC5, respectively.

The two-dimensional maps of Fig.~\ref{fig:prediction} represent
so-called \emph{biplots}, because they also feature the directions and
magnitudes of the PC loadings via the displayed arrows. Intuitively,
each arrow points at the correlation between a select two- or
three-body interaction and a PC. Fig.~\ref{fig:prediction} focuses on
two-body interactions only, and clearly highlights how the
\textsf{P4--Q0} and \textsf{Q0--Nda} both align perpendicular to the
separation between poor and high solute selectivity. The biplot
further singles out the role of \textsf{Q0} as a key solute bead to
modulate selectivity, but specifically identifies the bead's impact in
terms of two-body interactions.

Now that we have a two-dimensional map charted with clear basins of
poor and high selectivity, we apply it to predict the selectivity of
new solutes. We construct three compounds outside of the initial set
of $n=439$. Compounds A and B follow the three-dimensional structural
aspects prescribed by Figs.~\ref{fig:graphs}a and \ref{fig:3d_illu}a,
i.e., they are expected to be selective to CL. Compound C, on the
other hand, was originally eliminated from our initial study because
of a lack of stable insertion at the membrane interface, i.e.,
expected to \emph{not} be selective to CL.\cite{mohr2022data} For
these three compounds, we have no free-energy calculation to determine
$\Delta \Delta G$. We will instead solely apply the methodology
sketched in Fig.~\ref{fig:pipeline}: run a single MD simulation,
compute the structural order-parameter vector across the trajectory,
transform $\Delta\langle\bm{\mathcal X}\rangle$ to PCA coordinates,
and place the new compound on the two-dimensional PC3--PC5 map.

Fig.~\ref{fig:prediction}b  places the three
compounds A, B, and C on the two-dimensional map. C is featured well
within the basin of poorly selective compounds, where its lack of charged
beads places it toward the lower-right side of the map.
Compounds A and B, on the other hand, stand to the upper left of the
dividing line between poor and high selectivity, suggesting
selectivity to CL. For both compounds, the presence of charged beads
places them toward the leftmost side of PC3, while the number of
\textsf{T3} beads likely impact the different positions along PC5.
Naturally, a larger set of compounds would enrich the chemical space
explored, but even within our limits, we achieve reasonably accurate
predictions.

Evidently, estimating selectivity by transforming the solute's
$\Delta\langle\bm{\mathcal X}\rangle$ to PCA coordinates offers
significant appeal in terms of computational load. The alchemical
free-energy calculations involved in calculating $\Delta \Delta G$
consumed from 24 to 48 GPU hours of an NVIDIA Tesla V100 per neural
and charged compounds, respectively.\cite{mohr2022data} On the other
hand, a single MD simulation used in the present protocol only needed
0.3 to 0.7 GPU hours of an NVIDIA GTX 980. Though the two GPUs are
different, the need for a sole MD simulation---and without the usual
sensitivities associated with alchemical free-energy
calculations---evidently lead to a drastic reduction in computational
load.

\section{Conclusion}

The present work proposes a methodology based on molecular simulations
to link chemical structure to thermodynamic properties. Attempting a
direct structure--property link, e.g., via machine learning, between
chemical compound and target property is likely to be clouded by
several factors. First, the condensed-phase environment of a liquid
will likely lead to a combination of covalent and non-covalent
interactions, and both may critically impact the target property. In
addition, a single three-dimensional molecular configuration is
unlikely to be representative, because of the phase-space (Boltzmann)
averaging inherent to thermodynamic quantities.

To address this challenge, we propose the use of an atomic
representation originally developed for machine learning of electronic
properties: the Spectrum of London Axilrod-Teller-Muto (SLATM) representation.
SLATM decomposes a configuration into a collection of increasing
\emph{body-order} interactions: single particle (one-body); pairwise
(two-body); and triplets (three-body). For each term, SLATM builds a
histogram of population of these interactions. The pairwise term is
reminiscent of the radial distribution function, which hints at the
adequacy of the representation for molecular liquids. We adapt SLATM
to average over snapshots of an isothermal-isobaric MD simulation, acting as a
proxy for a Boltzmann average. This adapted ensemble-SLATM
representation thereby addresses the two above-mentioned issues: ($i$)
it does not distinguish between covalent and non-covalent
interactions; and ($ii$) offers phase-space averaging.

We argue that this adapted ensemble-SLATM representation is
particularly amenable to establishing structure--property
relationships of thermodynamic properties. As application, we focus on
a complex biomolecular system: small molecules targeting (phospholipid)
cardiolipin (CL) membrane environments. We rely on a coarse-grained
(CG) resolution, not only for computational efficiency, but mostly for
its ability to reduce the size of chemical space, and thereby screen
across compounds more efficiently. The CG resolution allows us to
screen across a large subset of small drug-like molecules with
relatively few CG molecular structures. Though based on the
biomolecular CG Martini model, our solute compounds are represented
via a further reduced force field that defines fewer bead types.

Establishing here the structure--property map boils down to reducing
the dimensionality of the SLATM vectors. To demonstrate the benefits
of including relevant physics in the representation (e.g., phase-space
averaging or key two- and three-body interactions) we apply a simple,
linear statistical method: principal component analysis (PCA).
Transformation of the original coordinates to the main principal
components allows us to focus on a handful of dimensions, thereby
significantly reducing the dimensionality of the problem.

Our analysis shows that we can correlate the first main principal
components (PCs) against relevant physicochemical descriptors, as well
as CL selectivity---the target property itself---via a \emph{single}
PC. The linearity of PCA makes it possible to transform back from PCA
to SLATM coordinates in order to identify key two- and three-body
interactions that impact the various PCs. We isolate key CG bead types present
in higher-order interactions that overwhelmingly impact CL selectivity. In the
present case, this includes CG types \textsf{Q0}, \textsf{T3}, \textsf{T4},
and \textsf{T5}, interacting favorably with the \textsf{Nda} bead type on
CL's head group. The results offer direct prescriptions on the design
of solutes selective to CL.

Finally, we gain further insight by charting a two-dimensional map in
the PCA coordinates. A simple evaluation of all pairs of PCs reveals
one that surprisingly separates two clear basins of compounds: poor
and high CL selectivity. From this map it is straightforward to
predict a compound's thermodynamic CL selectivity based on its PCA
coordinates. Computationally, this methodology only requires a
(relatively short) MD simulation, as compared to expensive alchemical
free-energy calculations. We demonstrate the idea on three test
compounds out of the initial training set.

Though demonstrated on a CG model applied to CL-membrane selectivity,
we foresee the methodology to be generally applicable to molecular
simulations of a variety of thermodynamic properties.

\begin{acknowledgements}
We sincerely thank Ioana Ilie, Joseph Rudzinski, and Jocelyne Vreede
for critical reading of the manuscript. We acknowledge support from
the Sectorplan B\`eta \& Techniek of the Dutch Government. This work
was completed in part with resources provided by the Dutch national
e-infrastructure with the support of SURF Cooperative. Icons on Fig. 1
and TOC from \href{https://www.flaticon.com/}{flaticon.com}.
\end{acknowledgements}

\section*{Supporting Information Available}
 The supporting information is available free of charge at DOI \dots. All
code and data needed to reproduce the results can be found on
zenodo.\cite{zenodo}
\begin{itemize}
  \item \emph{Supporting Information: Condensed-phase molecular
representation to link structure and thermodynamics in molecular
dynamics}: Coarse-grained force field; additional technical
information on the preparation of the samples; full set of plots and
illustrations generated during the analysis and interpretation of PCA.
  \item \emph{Zenodo archive}: MD-trajectories and simulation
parameter files; SLATM representations; codes for generating the SLATM
representations, performing PCA and analysis of the
results.\cite{zenodo}
 \item \emph{GitHub repository}: Codes for generating the SLATM
representations, performing PCA and interpretation of the
PCA.\cite{Mohr_STRUCTURAL_ANALYSIS_2023}
\end{itemize}


%
%
%
%
\section*{TOC Graphic}
\includegraphics[width=0.9\linewidth]{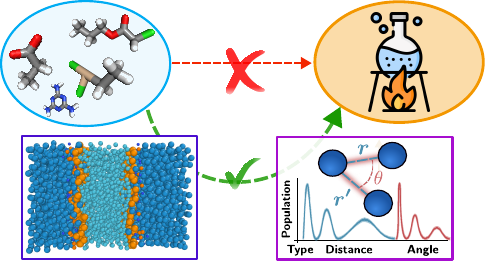}
%

\bibliography{structural_analysis.bib}

\end{document}